\documentclass[twocolumn,prl]{revtex4}
\usepackage{graphicx}
\usepackage{dcolumn}
\usepackage{bm}
\usepackage{amsmath}

\begin{document}

\preprint{}
\title{Theory of pairing symmetry inside the Abrikosov vortex core}
\author{Takehito Yokoyama$^1$,  Yukio Tanaka$^1$ and Alexander A. Golubov$^2$}
\affiliation{$^1$ Department of Applied Physics, Nagoya University, Nagoya, 464-8603, Japan and CREST%
\\
%and CREST, Japan Science and Technology Corporation (JST) Nagoya, 464-8603,
%Japan\\
$^2$ Faculty of Science and Technology, University of Twente, 7500 AE, Enschede, The Netherlands}
\date{\today}

\begin{abstract}
We show that the Cooper pair wave function at the center of an Abrikosov vortex with vorticity $m$
has different parity with respect to frequency from that in the bulk if $m$ is an odd number
and has the same parity if $m$ is an even number.
As a result, in a conventional vortex with $m=1$, the local density of states at the Fermi energy has a maximum (minimum)
at the center of the vortex core in even(odd)-frequency superconductor.
We propose a scanning tunneling microscope experiment using a superconducting tip to explore odd-frequency superconductivity.

%for a vortex with vorticity $m$ in
%superconductor, anomalous Green's function
%at the vortex center has the opposite (same) parity and frequency from (as) tho%se of the bulk if $m$ is an odd (even) integer.

%As examples, we study pairing states in even-frequency
%spin-singlet and odd-frequency
%spin-triplet even-parity $s$-wave superconductors and show that only
%odd and even-frequency spin-singlet chiral $p$-wave pairing is allowed at the
%center of the core, respectively, as a consequence of the broken translational
%symmetry. This is reflected by density of states inside the core.  Based on the%se results, we propose experimental setup to verify the
%existence of odd-frequency pairing in bulk materials by using scanning tunnelin%g spectroscopy.
\end{abstract}

\maketitle

%--- title ---
%--- author ---
%
%--- address ---
%
%--- date ---
%   Abstract
%-----------------------------------------------------------

%-----------------------------------------------------------

%\section{Introduction}

The study of the mixed state in type-II superconductors has
a long history and revealed a variety of physical
phenomena.\cite{Abrikosov}
In the clean limit,
low-energy bound states (the Andreev bound states) are generated in
the vortex core due to the spatial structure of the superconducting
pair potential. \cite{Caroli,Hess} One of
the manifestations of the bound states is the enhancement of
zero-energy quasiparticle density of states (DOS) locally in the
core, observable as a zero-bias conductance peak by scanning
tunneling microscope (STM).\cite{Hess,Fischer}
However, despite extensive studies of the vortex core, the issue of
pairing symmetry in the core remains unexplored.

Generally, superconducting pairing is classified into
even-frequency or odd-frequency state according to a symmetry with
respect to time. Due to the Fermi statistics, even-frequency
superconductors belong to the symmetry class of spin-singlet even-parity
(ESE) or spin-triplet odd-parity (ETO) pairing state, while odd-frequency
superconductors belong to the spin-singlet odd-parity (OSO) or
spin-triplet even-parity (OTE) pairing state.
Although vortex core state of even-frequency superconductors
has been well studied, that in odd-frequency superconductors
has not been clarified yet.

The possibility of the odd-frequency pairing state in various kinds
of uniform systems was discussed in Refs.
\cite{Berezinskii,Fuseya}, albeit its realization in
bulk materials is still controversial.
%It is really necessary to
%propse an idea to explore such unknown superconductor.
On the other hand, the
realization of the odd-frequency pairing state in inhomogeneous
even-frequency superconducting systems has recently been proposed. It is
established that odd-frequency pairing is induced due to symmetry
breaking in such systems. In ferromagnet/superconductor junctions,
odd-frequency pairing emerges due to the broken symmetry in a spin
space.\cite{Efetov1} It was recently realized that
proximity-induced odd-frequency pairing may be generated near
normal metal/superconductor interfaces due to the breakdown of
translational symmetry \cite{Tanaka3} or in a diffusive normal metal
attached to a spin-triplet superconductor.\cite{Tanaka2006}

Since an Abrikosov vortex breaks translational
symmetry in a superconductor, one may expect the emergence of an
odd-frequency pairing state around the vortex core even in a conventional
spin-singlet $s$-wave superconductor. On the other hand, one may imagine that
even-frequency pairing state may be generated around
the vortex core in odd-frequency superconductor.
%This significant difference may be useful to explore an odd-frequency superconductor \cite{Berezinskii,Fuseya}.
%

In this Letter, based on the quasiclassical theory of superconductivity, we
develop a general theory of pairing symmetry in an Abrikosov vortex core in clean superconductors, including odd-frequency superconductivity.
We show that for a vortex with vorticity $m$ in
superconductor, pairing function of the Cooper pair
at the vortex center has the opposite (same) symmetry with respect to frequency  to (as) that of the bulk if $m$ is an odd (even) integer.
For a conventional vortex with $m=1$, we show that
zero energy local DOS is enhanced (suppressed) at the
center of the vortex core in even (odd)-frequency superconductor.
We further reveal that OSO $p$-wave pairing is generated at the
center of the core of ESE $s$-wave superconductor.
On the other hand, in OSO $p$-wave superconductor,
ESE $s$-wave pairing state emerges at the center of the vortex core.
%As examples, we study pairing states in ESE $s$-wave and
%OTE $s$-wave superconductors with $m=1$ and show that only
%odd and even-frequency spin-singlet chiral $p$-wave pairing is allowed at the
%center of the core, respectively.
Based on these results, we propose an experimental setup to explore odd-frequency superconductivity by probing a local Josephson coupling by STM with superconducting tip.

%If we use ESE $s$-wave superconductor as a tip,
%Josephson current is absent (nonzero)
%between tip and vortex core (bulk) for ESE $s$-wave case.
%On the other hand, it is nonzero (absent) for OSO $p$-wave case. \par

%It is well known that the sign change occurs on a trajectory passing near the c%enter of a vortex and therefore leads
%to localized Andreev bound states inside the vortex core.
%We will show that this ZEP is a direct consequence of odd frequency pairing.

The electronic structure of the vortex core in a single Abrikosov vortex in a
clean superconductor is described by the quasiclassical Eilenberger
equations \cite{Eilenberger,Larkin}
%
%For the calculation of the
%Green's functions in the vicinity of the vortex, it is necessary to
%find numerically stable solutions of the Eilenberger equations. For
%this purpose, we use the
based on the Riccati parametrization of the
quasiclassical propagator \cite{SchopohlMaki}. Along a trajectory
${\bm r}(x') = {\bm r}_0 + x' \; \hat{\bm {v}}_F $ with unit vector
$ \hat{\bm v}_F$  parallel to $\bm {v}_F$, the Eilenberger equations
are generally represented in 4$\times4$ matrix form.\cite{Eschrig}
For a singlet (triplet) superconductor with $\hat{\Delta}=\Delta \sigma_y (\sigma_x)$ 
($\sigma_x$ and $\sigma_y$ are  Pauli's matrices in spin space)\cite{note}, 
these equations are reduced to the set of two decoupled differential equations of the Riccati type for the functions $a(x')$ and $b(x')$,
\begin{eqnarray}
\hbar v_F \partial_{x'}  a(x') + \left[ 2 \epsilon_n + \Delta^\dagger a(x') \right] a(x')
- \Delta & = & 0, \nonumber \\
\hbar v_F \partial_{x'} b(x') - \left[ 2 \epsilon_n + \Delta b(x') \right] b(x') + \Delta^\dagger & = & 0 \label{Riccati}
\end{eqnarray}
where $ i \epsilon_n$ are the Matsubara frequencies, and $\Delta^\dagger=(-)\Delta^*$ for even (odd)-frequency superconductor. For simple case
of a cylindrical Fermi surface, the Fermi velocity can be written as
%\begin{equation}
${\bm v}_F = v_F ({\bm e}_1 \cos \theta + {\bm e}_2 \sin \theta).$
%\end{equation}
%The $\theta$- and ${\bm r}$-dependence of the pairing potential $\Delta$ can be% factorized
%in the form

We choose the following form of the pair potential:
%which
%describes the behavior around the vortex:
%\begin{eqnarray}
% \Delta ({\bf{r}},\theta,E) &
%=& \Delta _0 \chi(\theta,E) \tanh \left( {\frac{{\sqrt {x^2  + y^2 } }}{\xi }} %\right) \left( {\frac{{x + iy}}{{\sqrt {x^2  + y^2 } }}} \right)^m  \nonumber \%\
% & \equiv & \Delta _0 \chi(\theta,E) \tanh \left( {\frac{{\sqrt {x^2  + y^2 } }%}{\xi }} \right)\exp (im\varphi )  .
%\end{eqnarray}
\begin{equation}
\Delta ({\bf r},\theta,E)
%=& \Delta _0 \chi(\theta,E) \tanh \left( {\frac{{\sqrt {x^2  + y^2 } }}{\xi }} %\right) \left( {\frac{{x + iy}}{{\sqrt {x^2  + y^2 } }}} \right)^m  \nonumber \%\
=  \Delta _0 \chi(\theta,E) F(r) \exp (im\varphi )
\end{equation}
with $r=\sqrt {x^2  + y^2 }$ and $\exp(i\varphi)=(x + iy)/\sqrt {x^2  + y^2 }$.
Here,  $F(r)$ denotes the spatial profile of the gap, $m$ is the vorticity, and $\chi(\theta,E)$ is the symmetry function. Also, we introduce the coherence length $\xi = \hbar v_F/\Delta_0$,
the center of a vortex is situated at $x=y=0$, and $\exp(im\varphi)$ is
the phase factor which originates from the vortex.

%%For $\epsilon_n > 0$ the Riccati equations have to be solved using
%%the bulk values as initial values at $x' = \pm \infty$.
%\begin{eqnarray}
% a(-\infty) = \frac{\Delta (-\infty)}{\epsilon_n + \sqrt{\epsilon_n^2 + |\Delta
% (-\infty)|^2}}, \nonumber \\
% b(+\infty) = \frac{\Delta^\dagger (+\infty)}
%{\epsilon_n + \sqrt{\epsilon_n^2 +% |\Delta
% (+\infty)|^2}}.
%\end{eqnarray}
%Then,
We obtain pairing function of the Cooper pair (anomalous Green's function)
$f$ as $f = - 2a/(1 + ab)$.
%\begin{eqnarray}
%f = - \frac{2a}{1 + ab}.
%\end{eqnarray}
For the calculation of the local
DOS normalized by its value in the normal state,
 the quasiclassical propagator has to be integrated over the
angle $\theta$ which defines the direction of the Fermi velocity.
The normalized local DOS in terms of functions $a$ and $b$ is given by
\begin{equation}
N({\bm r}_0,E) = \int_0^{2\pi} \frac{d \theta}{2 \pi}  {\rm Re} \;
\left[ \frac{1-a  b}{1+ a b} \right]_{i \epsilon_n \rightarrow E + i
\delta}, \label{angularaverage}
\end{equation}
where $E$ denotes the quasiparticle energy with respect to the Fermi
level and $\delta$ is an effective scattering parameter that
corresponds to an inverse mean free path. In numerical calculations
throughout this paper, we will fix this value as
$\delta=0.1\Delta_0$.
% Further, in what follows, the origin $\mathcal{O}$ of our coordinate system is placed at the center of the vortex core.

%%%%%%%%%%%%%%%%%%%%%%%%%%%%%%%%
%\section{Results}
First, we discuss the general property of the
symmetry at the vortex center.
Vorticity and the symmetry of superconductor with respect to frequency crucially affect the symmetry of the Cooper pair
at the core center. Consider a trajectory passing through the center of the vortex. 
By setting $x'=0$ at the vortex center, we get $b(x',\epsilon _n ) =  - 1/a( - x', - \epsilon _n )$ from the Eilenberger equations for even-frequency superconductor with odd integer $m$ or odd-frequency superconductor with even integer $m$. Similarly, we obtain $b(x',\epsilon _n ) =  1/a( - x', - \epsilon _n )$ for odd-frequency superconductor with odd integer $m$ or even-frequency superconductor with even integer $m$. Thus, at the vortex center $x'=0$,  we get $ f(\epsilon _n ) =  - f( - \epsilon _n )$ in the former case, while $ f(\epsilon _n ) =   f( - \epsilon _n )$ in the latter. 
 Note that spin is conserved in the vortex state considered. Therefore, quite generally, for an odd  integer $m$ the induced pairing 
at the vortex center has different symmetry with respect to frequency from that in the bulk 
superconductor. On the other hand, for an even integer $m$, the induced pairing  at the vortex center has the same symmetry as that of the bulk. We summarize
pairing symmetry at the vortex center in Table I.
%Then odd-frequency pairing state is generated when $m$ is an odd
%number. Then the local DOS has a ZEP.
%On the other hand, for an even integer $m$, the corresponding
%zero energy peak is absent.

For conventional $s$-wave case, there have been
several studies of multi-vortex state with $m \ge 1$
\cite{multi,SSC}. It was shown that zero energy peak in the DOS only appears
for odd number $m$ at the vortex center \cite{SSC}. This statement is consistent with our result for the conventional $s$-wave case of $\chi(\theta,E)=1$ because odd-frequency pairing state is generated only for odd integer $m$.
The relation between zero energy peak in DOS and odd-frequency pairing state will be discussed later.

\begin{center}
\begin{table}[h]
\caption{Pairing symmetry in the vortex state. }
\begin{tabular}{|c|p{1.4cm}|p{1.4cm}|p{1.9cm}|p{1.9cm}|}
\hline
& bulk state & vorticity  m  & symmetry at the center \\ \hline
(1) & ESE  & odd  & OSO \\ \hline
(2) & ESE  & even  & ESE \\ \hline
(3) & ETO  & odd  & OTE \\ \hline
(4) & ETO & even  & ETO \\ \hline
(5) & OSO & odd  & ESE \\ \hline
(6) & OSO & even  & OSO \\ \hline
(7) & OTE & odd  & ETO \\ \hline
(8) & OTE & even  & OTE \\ \hline
\end{tabular}%
\end{table}
\end{center}

\begin{figure}[htb]
\begin{center}
\scalebox{0.4}{
\includegraphics[width=17.5cm,clip]{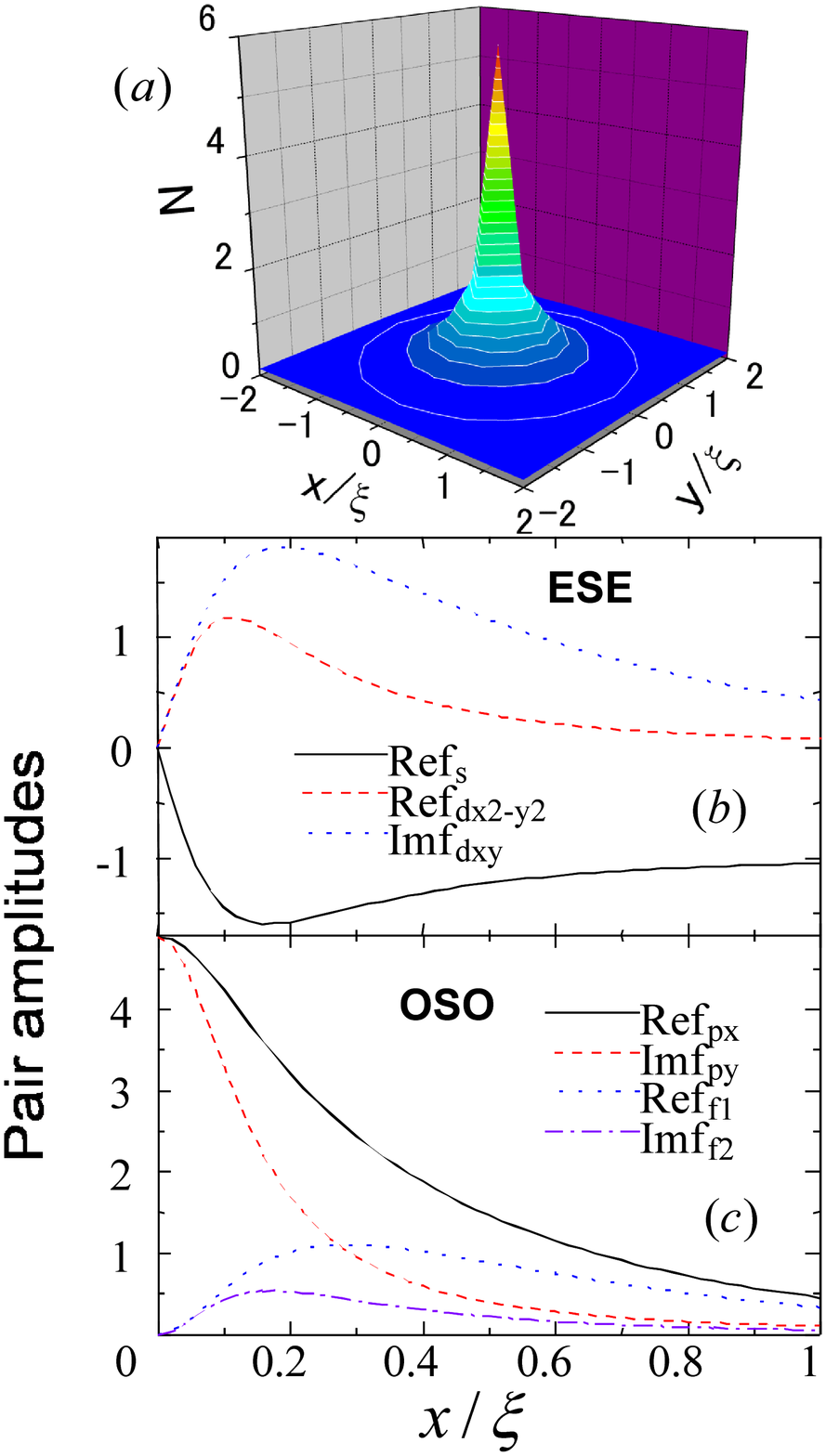}}
\end{center}
\caption{(color online) Results for ESE $s$-wave superconductor. (a) Normalized local DOS around the vortex at $E=0$. The center
of the vortex is situated at $x=y=0$. Spatial dependencies of (b) ESE and (c) OSO components at $E=0$.
Only OSO components, Re$f_{px}$ and
Im$f_{py}$, can survive inside the core (near $x=0$).} \label{f1}
\end{figure}

\begin{figure}[htb]
\begin{center}
\scalebox{0.4}{
\includegraphics[width=17.5cm,clip]{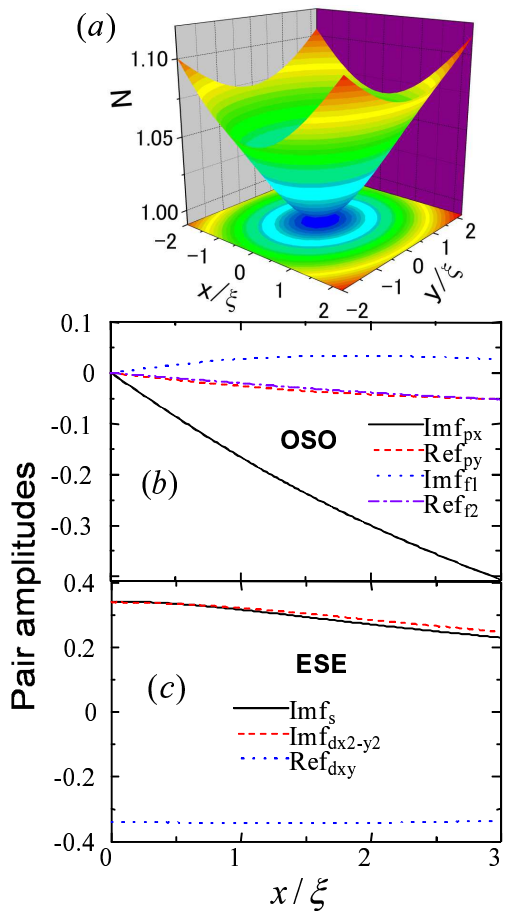}}
\end{center}
\caption{(color online) Results for OSO $p_x$-wave superconductor.
 (a) Normalized local DOS around the vortex at $E=0$. The center of the vortex is situated at $x=y=0$.  Spatial dependencies of (b) OSO and (c) ESE components at $E=0$. Only ESE components can exist inside the core (near $x=0$).} \label{f2}
\end{figure}

%\begin{figure}[htb]
%\begin{center}
%\scalebox{0.4}{
%\includegraphics[width=21.0cm,clip]{fig3.eps}}
%\end{center}
%\caption{(color online) $p$-wave pairing components at $x=y=0$ as a function of energy $E$. The relations ${\mathop{\rm Re}\nolimits} f_{px}  = {\mathop{\rm Im}\nolimits} f_{py}$ and ${\mathop{\rm Im}\nolimits} f_{px}  =  - {\mathop{\rm Re}\nolimits} f_{py}$ hold for all $E$.} \label{f3}
%\end{figure}

%\begin{figure}[htb]
%\begin{center}
%\scalebox{0.4}{
%\includegraphics[width=19.0cm,clip]{fig3.eps}}
%\end{center}
%\caption{(color online) Suggested experimental setup to verify the
%existence of odd-frequency pairing in bulk materials. Nonzero local
%Josephson current is allowed at the core when superconducting STM
%tip has odd-frequency spin-singlet odd-parity $p$-wave pairing.
%Here, $r=\sqrt {x^2  + y^2 }$. } \label{f3}
%\end{figure}

In general, the most realizable vorticity is $m=1$.
Thus, in the following, we will study in detail two typical cases at $m=1$ with ESE $s$-wave and OSO $p_x$-wave superconductors where we choose $\chi(\theta,E)=1$ and $\chi(\theta,E)=(C \cos\theta E/\Delta _0) /[1 + \left( {E/\Delta _0 } \right)^2 ]$ with $C=0.8$,\cite{Tanaka2006} respectively. Also, spatial dependence of the gap is chosen as $F(r)=\tanh (r/\xi)$. 

Due to the broken translational symmetry of the system,
various pairing states are
expected to emerge around the vortex. In order to study 
possible pairing states, we decompose anomalous
Green's function $f$ into various angular momentum components as
follows
\begin{eqnarray}
f \cong f_s  + f_{px} \cos \theta  + f_{py} \sin \theta  + f_{dx^2  - y^2 } \cos 2\theta  \nonumber \\
 + f_{dxy} \sin 2\theta + f_{f1 } \cos 3\theta  + f_{f2} \sin 3\theta  + .....
\end{eqnarray}
Note that \textit{ all the above pairing components are spin singlet}.

Figure \ref{f1} shows the results for ESE $s$-wave superconductor.
The local DOS around the vortex at $E=0$ is shown in Fig. \ref{f1} (a). As is well known, zero energy peak appears in the core \cite{Hess}.
The spatial dependencies of decomposed anomalous
Green's function $f$ at $E=0$ are shown in Figs. \ref{f1} (b) and (c).
Interestingly, only OSO pairing components,
Re$f_{px}$ and Im$f_{py}$, survive at the center of the core. With
the increase of the distance from the core center, the magnitudes of
$f$ decrease rapidly, except for $s$-wave
one. Note that other angular momentum components not shown in this
figure are negligibly small.
We also find that the following representation of anomalous Green's function $f$ holds at the center of the core:
\begin{eqnarray}
f = \left( {{\mathop{\rm Re}\nolimits} f_{px}  + i{\mathop{\rm Im}\nolimits} f_{px} } \right) (\cos\theta + i\sin\theta ) = f_{px} \exp (i\theta ).
\end{eqnarray}
Thus, we see that anomalous Green's function at the core center has chiral $p$-wave symmetry.

%We see that anomalous Green's function at the core center has the same angular %dependence as the gap function.

%Figure \ref{f2} shows the decomposition of the $p$-wave pair amplitude at the center of the core $x=y=0$ as a function of energy. Note that other pairing amplitudes are absent at $x=y=0$ for all $E$. The relations  ${\mathop{\rm Re}\nolimits} f_{px} = {\mathop{\rm Im}\nolimits} f_{py}$ and ${\mathop{\rm Im}\nolimits} f_{px}  =  - {\mathop{\rm Re}\nolimits} f_{py}$ are satisfied for all $E$, as illustrated on the figure.

The enhancement of the local DOS in the presence of odd-frequency
pairing can be understood, irrespective of the detailed
shape of $\Delta$ by using the normalization condition for
the quasiclassical Green's functions, $g^2+f\bar f=1$. Indeed, since for odd-frequency pairing state, the anomalous Green's function $\bar f
=- 2b/(1+ab)$ at $E=0$ is given by $\bar f(\theta ) =  - f^*
(\theta)$ (see Ref.\cite{Eschrig}) and local DOS is given by
$N(E)=-$Re$g$, one can show that generally $N(E=0)>1$ since $g^2=1+\left| f \right|^2 >1$. This means
that the emergence of the odd-frequency pairing is a physical reason
of zero energy peak of the local DOS inside the core.
The manifestation of the odd-frequency chiral $p$-wave pairing state
at the center of the vortex core
%The above discussions makes it possible to provide new interpretation of
%the Andreev bound states inside the core in ESE $s$-wave superconductors as
%\textit{the manifestation of the odd-frequency pairing}. This interpretation
is also consistent with the experimental fact that
the observed zero-bias conductance peak by STM at a vortex center
is very sensitive to disorder \cite{Renner}, since $p$-wave pairing is
fragile to impurity scattering. 

Figure \ref{f2} depicts the results for OSO $p_x$-wave superconductor.
The local DOS around the vortex at $E=0$ is shown in Fig. \ref{f2} (a). 
In dramatic contrast to the result for the ESE $s$-wave, zero energy DOS is suppressed at the core.  
%This is reflected by the pairing symmetry of even frequency at the core.
The spatial dependencies of decomposed anomalous
Green's function at $E=0$ are shown in Figs. \ref{f2} (b) and (c).
As is seen, only ESE pairing components exist at the center of the core. 
For even-frequency pairing state, $\bar f(\theta ) = f^* (\theta)$ is satisfied
at $E=0$ and hence we get $N(E=0)<1$ which is consistent with Fig. \ref{f2} (a).
%as a chiral $p$-wave state, similar to the results of the ESE $s$-wave case.
%With the increase of the distance from the core center, the magnitudes of
%anomalous Green's functions decrease, except for $s$-wave
%one.
By comparing Figs.1 and 2, it is  clear that
zero energy local DOS $N(0)$ has a maximum at the
center of the vortex core in even-frequency superconductor, while it
has a minimum at the core center in odd-frequency superconductor.
This  difference can be detected by STM. 

As regards the candidate for the odd-frequency superconductor,
CeCu$_2$Si$_2$ and CeRhIn$_5$ are possible materials
\cite{Fuseya,Kawasaki}.
In these systems, OSO state with $p$-wave symmetry is considered to be
promising \cite{Fuseya}. In the light of the present theory,
ESE $s$-wave pairing is expected to appear inside the vortex core.
Based on this idea, we propose an experimental setup to verify the existence of
odd-frequency pairing in bulk materials by using  superconducting
STM where we use conventional $s$-wave superconductor as a STM tip\cite{Kohen}
as shown in Fig. 3.
\begin{figure}[htb]
\begin{center}
\scalebox{0.4}{
\includegraphics[width=15.0cm,clip]{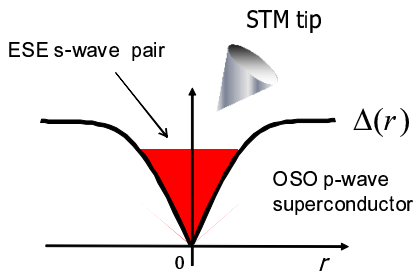}}
\end{center}
\caption{(color online) Suggested experimental setup to probe the
local Josephson coupling between
superconducting tip and superconductor
with Abrikosov vortex core.  } \label{f4}
\end{figure}
%\cite{Hess,Fischer,Kohen} As discussed above, ESE $s$-wave pairing state is all%owed at the center of the vortex core in OSO $p$-wave superconductor. As a resu%lt, a gap structure of the DOS at the core center is expected.
%Therefore, using a normal STM tip,\cite{Hess,Fischer} one can study the DOS and% hence the pairing symmetry at the core center. Then, if one finds a gap struct%ure of the DOS, it strongly suggests the existence of bulk OSO $p$-wave superco%nductor.
%Also, using an ESE $s$-wave superconductor as an STM tip\cite{Kohen} and
Local Josephson current measured in
STM experiment with superconducting tip is given by \cite{Zaitsev}
\begin{eqnarray}
eIR  = \pi T\sum\limits_{\theta ,\epsilon _n } {{\mathop{\rm Im}\nolimits} \left( {f^* (\theta ,\epsilon _n )f_S (\theta ,\epsilon _n )} \right)}.
\end{eqnarray}
Here, $R$ is the junction resistance, $T$ is temperature, and $f_S$
is anomalous Green's function in the STM tip.
It follows from this
expression that finite Josephson current is allowed only when
superconducting STM has the same symmetry as that in the vortex
state. \textit{Therefore,
finite Josephson current is allowed at the vortex core only when the pairing symmetry at the core is ESE s-wave}.
First, we consider bulk superconductor which has an ESE $s$-wave symmetry with $m$=1.
As seen from Fig. \ref{f1}, only OSO state is generated at the core center.
Then, the local Josephson coupling is absent at the
core while it exists in the bulk.
On the other hand, in OSO $p$-wave superconductor with $m=1$, 
ESE $s$-wave state is generated inside the core (see Fig. \ref{f2}).
In this case, the local Josephson coupling only exists at the 
core. Such a dramatic feature is useful to explore and identify odd-frequency 
pairing in bulk superconductor. We summarize the possible local Josephson couplings with ESE $s$-wave STM tip in Table II.

\begin{center}
\begin{table}[h]
\caption{Local Josephson coupling with ESE $s$-wave STM tip.}
\begin{tabular}{|c|c|c|}
\hline
 & \multicolumn{2}{c|}{Local Josephson coupling}
\\ \cline{2-3}
Bulk state & \hspace{0.5cm} core \hspace{0.50cm} & bulk \\ \hline
ESE & No &  Yes \\ \hline
OSO & Yes&  No \\ \hline
\end{tabular}
\end{table}
\end{center}
%The experimental setup is illustrated in Fig. \ref{f3}.

%Finally, we propose an experimental setup to verify our finding. Our theory pre%dicts that only odd-frequency spin-triplet even-parity pairing (OTE) is allowed% at the center of the vortex core in chiral $p$-wave superconductor, realized i%n Sr$_2$RuO$_4$.\cite{Maeno,Mackenzie} It is established that OTE pairing is al%so generated in half-metal/$s$-wave superconductor junctions.\cite{Asano,Kaizer%} Then, using superconducting STM tip made of half-metal attached to an $s$-wav%e superconductor, one may confirm the existence of odd frequency pairing by obs%erving zero bias peak at the core. Note that if we use superconducting STM tip %made of conventional $s$-wave superconductor instead, we cannot observe a zero %bias peak.  We show the experimental setup in  Fig. \ref{f4}.

In summary, we have developed a general theory of pairing symmetry
inside the Abrikosov vortex core in superconductors, including odd-frequency superconductivity. 
We have found that for a vortex with vorticity $m$ in a
superconductor, anomalous Green's function
at the vortex center has the opposite (same) symmetry with respect to frequency  to (as) that of the bulk if $m$ is an odd (even) integer.
%As examples, we have studied pairing states in ESE $s$-wave and OTE $s$-wave su%perconductors and shown that only
%odd and even-frequency spin-singlet chiral $p$-wave pairing is allowed at the
%center of the core, respectively, as a consequence of the broken translational
%symmetry.
We have also shown that
zero energy local density of states is enhanced (suppressed) at the
center of the vortex core for even(odd)-frequency superconductor.
Based on the obtained results, we proposed a
scanning tunneling microscope experiment using a superconducting tip to detect
local Josephson coupling in order to explore and identify  odd-frequency superconductor.
% Based on these
%results, we have proposed an experimental setup to explore the
%odd-frequency pairing in bulk materials by using STM.

%The relation between odd-frequency pairing and zero energy peak of local DOS is quite general feature as already found in normal metal/superconductor junctions\cite{Tanaka3,Tanaka2006} or ferromagnet/superconductor junctions.\cite{Braude} The interpretation of Abrikosov vortex as {\it a manifestation of the odd-frequency pairs} may become a useful concept to explore new features in the vortex physics.

%%%%%%%%%%%%%%%%%%%%%%%%%%%%%%%%%%%%%%%%%%%%%%%%%%%
%
%==================================================
We acknowledge valuable discussion with C. Iniotakis, Y. Asano and
N. Hayashi.
T.Y. acknowledges support by the JSPS.
This work was supported by Ministry of Education, Culture,
Sports, Science and Technology, Grant No. 17071007 and NanoNed Grant
TSC7029.

%======Reference===================================
%

%===============================================================

\begin{thebibliography}{99}
\bibitem{Abrikosov}
A. A. Abrikosov, Superconductivity, edited by R.D. Parks
(Marcel Dekker, New York, 1969);
G. Blatter, M. V. Feigel'man, V. B. Geshkenbein, A. I. Larkin, and V. M. Vinokur, Rev. Mod. Phys. {\bf 66}, 1125 (1994);
T. Maniv, V. Zhuravlev, I. Vagner, and P. Wyder,
Rev. Mod. Phys. {\bf 73}, 867 (2001).

\bibitem{Caroli} C. Caroli, P. G. de Gennes and J. Matricon, Phys. Lett. \textbf{9}, 307 (1964); Yu. G. Makhlin and G. E. Volovik, JETP Lett. \textbf{62}, 737 (1995); A. I. Larkin and Yu. N. Ovchinnikov, Phys. Rev. B \textbf{57}, 5457 (1998); N. B. Kopnin and G. E. Volovik, Phys. Rev. Lett.
\textbf{79}, 1377 (1997); Phys. Rev. B \textbf{57}, 8526 (1998);
G. E. Volovik, JETP Lett. \textbf{70}, 609 (1999).

\bibitem{Hess} H. F. Hess, R. B. Robinson, R. C. Dynes, J. M. Valles, Jr., and J. V. Waszczak,
Phys. Rev. Lett. \textbf{62}, 214 (1989);
F. Gygi and M. Schl\"{u}ter, Phys. Rev. B 43, 7609 (1991).

\bibitem{Fischer} \O. Fischer, M. Kugler, I. Maggio-Aprile, C. Berthod, and C. Renner,
Rev. Mod. Phys. \textbf{79}, 353 (2007).


\bibitem{Berezinskii} V. L. Berezinskii, JETP Lett. \textbf{20}, 287 (1974);
A. Balatsky and E. Abrahams, Phys. Rev. B \textbf{45},
13125 (1992); E. Abrahams, A. Balatsky, D. J. Scalapino and J. R. Schrieffer,
Phys. Rev. B \textbf{52}, 1271 (1995); 
P. Coleman, E. Miranda, and A. Tsvelik, Phys. Rev. Lett. \textbf{70}, 2960
(1993); Phys. Rev. B \textbf{49}, 8955 (1994);
M. Vojta and E. Dagotto, Phys. Rev. B \textbf{59}, R713 (1999).

\bibitem{Fuseya} Y. Fuseya, H. Kohno and K. Miyake, J. Phys. Soc. Jpn.
\textbf{72}, 2914 (2003).

\bibitem{Efetov1} F. S. Bergeret, A. F. Volkov, and K. B. Efetov, Phys. Rev.
Lett. \textbf{86}, 4096 (2001); F. S. Bergeret, A. F. Volkov, and K. B. Efetov, Rev. Mod. Phys. \textbf{77}, 1321 (2005).

%\bibitem{Kadigro} A. Kadigrobov, R. I. Shekter and M. Jonson, EuroPhys. Lett. \textbf{90}, 394 (2001).

\bibitem{Tanaka3} Y. Tanaka, A. A. Golubov, S. Kashiwaya, and M. Ueda,
Phys. Rev. Lett. {\bf  99}, 037005 (2007);
M. Eschrig, T. Lofwander, Th. Champel, J. C. Cuevas and G. Schon,
J. Low Temp. Phys. \textbf{147} 457 (2007).
%Y. Tanaka, Y. Tanuma and
%A. A. Golubov, Phys. Rev. B {\bf 76}, 054522 (2007).

\bibitem{Tanaka2006} Y. Tanaka and A. A. Golubov, Phys. Rev. Lett. \textbf{98}, 037003 (2007).

%\bibitem{Fetter} A. L. Fetter and P. C. Hohenberg, in \textit{Superconductivity}, edited by R. D. Parks (Marcel Dekker, New York, 1969).

\bibitem{Eilenberger} G.~Eilenberger, Z.~Phys. {\bf 214}, 195 (1968).

\bibitem{Larkin} A.~I.~Larkin and Yu.~N.~Ovchinnikov, Sov.~Phys.~JETP {\bf 28}, 1200 (1969).

\bibitem{SchopohlMaki} N.~Schopohl and K.~Maki, {Phys.~Rev.~B} {\bf 52}, 490 (1995).
%N.~Schopohl, cond-mat/9804064 (unpublished).

\bibitem{Eschrig} M. Eschrig, Phys. Rev. B {\bf 61}, 9061 (2000).

\bibitem{note} For triplet pairing, we can choose d-vector parallel to z-axis without loss of generality.

\bibitem{multi} G.~E.~Volovik, JETP Lett. {\bf 57}, 244 (1993); S.~M.~M.~Virtanen and M.~M.~Salomaa, Phys. Rev. B {\bf 60} 14581 (1999).

\bibitem{SSC}Y. Tanaka, H. Takayanagi and A. Hasegawa,
Solid state Commun. {\bf 85}, 321 (1993); A. S. Mel'nikov and V. M. Vinokur,
Nature {\bf 415}, 60 (2002).
%A.S. Mel'nikov and V.M. Vinokur,
%Phys. Rev. B {\bf 65} 224514 (2002).

\bibitem{Renner} C. Renner, A. D. Kent, P. Niedermann, O. Fischer, and F. L\'evy, Phys. Rev. Lett. {\bf 67}, 1650 (1991).

\bibitem{Kawasaki} S. Kawasaki, T. Mito, Y. Kawasaki, G.-q. Zheng, Y. Kitaoka, D. Aoki, Y. Haga, and Y. Onuki, 
Phys. Rev. Lett. {\bf 91}, 137001 (2003);
Guo-qing Zheng, N. Yamaguchi, H. Kan, Y. Kitaoka, J. L. Sarrao, P. G. Pagliuso, N. O. Moreno, and J. D. Thompson, Phys. Rev. B {\bf 70}, 014511 (2004). 

\bibitem{Kohen} A. Kohen, Th. Proslier, T. Cren, Y. Noat, W. Sacks, H. Berger, and D. Roditchev, Phys. Rev. Lett. {\bf 97}, 027001 (2006);
J.G. Rodrigo, H. Suderow, and S. Vieiraa, Eur. Phys. J. B {\bf 40},
483 (2004).

\bibitem{Zaitsev} A. V. Zaitsev, Sov. Phys. JETP {\bf 59}, 1015 (1984).


\end{thebibliography}
\end{document}